\documentclass[12pt]{article}
\usepackage{amsmath,amssymb}
\usepackage{graphicx}
\usepackage{color}
\usepackage{subfigure}

\begin{document}
\baselineskip=16pt
\begin{titlepage}
\begin{flushright}
{\small SU-HET-04-2013}

{\small EPHOU-13-009}

{\small IPMU13-0180}
\end{flushright}
\vspace*{1.2cm}

\begin{center}

{\Large\bf NNMSM Type-II and III} 
\lineskip .75em
\vskip 1.5cm

\normalsize
{\large Naoyuki Haba}$^{1,2}$,
{\large Kunio Kaneta}$^{2,3,4}$ and 
{\large Ryo Takahashi}$^2$

\vspace{1cm}
$^1${\it Graduate School of Science and Engineering, Shimane University, 

Matsue, Shimane 690-8504, Japan}\\
$^2${\it Department of Physics, 
Faculty of Science, Hokkaido University, 

Sapporo, Hokkaido 060-0810, Japan}\\
$^3${\it Kavli Institute for the Physics and Mathematics of the Universe (WPI), 
University of Tokyo, Kashiwa, Chiba 277-8568, Japan}\\
$^4${\it Department of Physics, Graduate School of Science, Osaka 
University, 

Toyonaka, Osaka 560-0043, Japan}\\

\vspace*{10mm}

{\bf Abstract}\\[5mm]
{\parbox{13cm}{\hspace{5mm}
We suggest two types of extension of the standard model, which are the so-called
 next to new minimal standard model (NNMSM) type-II and -III. They can achieve 
gauge coupling unification as well as suitable dark matter abundance, small 
neutrino masses, baryon asymmetry of the universe, inflation, and dark energy. 
The gauge coupling unification can be realized by introducing extra two or three
 new fields, and could explain the charge quantization. We also show that there 
are regions in which the vacuum stability, coupling perturbativity, and correct 
dark matter abundance can be realized with current experimental data at the same
 time.
}}

\end{center}

\end{titlepage}

\section{Introduction}
A discovery of the Higgs particle at the Large Hadron Collider (LHC) 
experiment~\cite{Chatrchyan:2013lba} filled the last piece of the standard model
 (SM). Furthermore, the results from the LHC experiment are almost consistent 
with the SM, and no signatures of the supersymmetry (SUSY) are discovered. 
However, there are some unsolved problems in the SM, e.g., the SM does not have 
a dark matter (DM) candidate although a SUSY model can give it. The SUSY is one 
of excellent candidates beyond the SM because it can also solve the gauge 
hierarchy problem in addition to the DM. Moreover, the gauge coupling 
unification (GCU) can be realized in a minimal supersymmetric standard model 
(MSSM). But, the discovery of the Higgs with the 126 GeV mass and no signatures 
of the SUSY may disfavor the SUSY at low energy. 

Actually, various extensions of the SM without the SUSY have been proposed. One 
minimal extension is the new minimal standard model 
(NMSM)~\cite{Davoudiasl:2004be}. The NMSM contains a gauge singlet real scalar, 
two right-handed neutrinos, an inflaton, and the small cosmological constant, 
which can explain the DM, small neutrino masses and baryon asymmetry of the 
universe (BAU), inflation, and dark energy (DE), respectively.\footnote{See 
also~\cite{Ibe:2009gt,Chpoi:2013wga,Marshak:1979fm,Asaka:2005an} for other 
extensions of the SM.} Then, the next to new minimal SM (NNMSM) has been 
suggested in Ref.\cite{hkt0}. The NNMSM adds new particle contents to the NMSM 
in order to explain the gauge coupling unification (GCU). These are two adjoint 
fermions and four vector-like fermions.\footnote{The model did not address the 
gauge hierarchy problem because the magnitude of the fine-tuning is much less 
than that of the cosmological constant problem.} In the NNMSM, the stability and
 triviality conditions have also been analyzed by use of recent experimental 
data of Higgs and top masses\cite{Beringer:1900zz}. As a result, it could be 
found that there are parameter regions in which the correct abundance of DM can 
also be realized at the same time. One is the lighter DM mass, $m_S$, region as 
$63.5~{\rm GeV}\lesssim m_S \lesssim 64.0~{\rm GeV}$, and the other is the 
heavier one as $708~{\rm GeV}\lesssim m_S \lesssim 2040~{\rm GeV}$ with the 
center value of top pole mass. 

Ref.\cite{hkt0} has introduced six new fields, which were assumed to be the same
 mass scale, in order to realize the GCU. In this work, we also extend the NMSM 
and leave the condition of the same mass scale for the new fields adopted in the
 NNMSM. As a result, we can reduce the particle contents of the NNMSM while 
realizing the GCU and vacuum stability, and satisfying phenomenological 
constraints such as DM, small neutrino masses, BAU, inflation, DE, and the 
proton decay. We suggest two types of model, the first model includes two 
adjoint fermions for the GCU and the second one has three adjoint fermions. The 
degrees of freedom of additional fermions in the models decrease compared to the
 NNMSM. We call the models as NNMSM type-II (two adjoint fermions) and NNMSM 
type-III (three adjoint fermions), respectively. The NNMSM type-II also includes
 two right-handed neutrinos like the NNMSM and the type-III does not have them. 
Both models has the gauge singlet scalar, inflaton, and the small cosmological 
constant. We also analyze the stability and triviality bounds with the 126 GeV 
Higgs mass, the recent updated limits on the DM particle\cite{Cline:2013gha}, 
and the latest experimental value of the top pole mass as 173.5 GeV in both 
models. We will point out that there are parameter regions in both models, where
 the stability and triviality bounds, the correct abundance of DM, and the Higgs
 and top masses can be realized at the same time. 

This paper is organized as follows: In Section 2, we suggest the NNMSM type-II 
and show that the model can realize the GCU. Then, the vacuum stability and DM 
are investigated. Some related arguments in the model such as the inflation, 
neutrino, and baryogenesis are also given. In Section 3, we suggest the NNMSM 
type-III. In particular, we focus on two simple models in this type of model. 
The similar phenomenological arguments to those in type-II are also presented. 
Section 4 is devoted to the summary. We give other setups for the realization of
 GCU in Appendix.

\section{NNMSM type-II}

\subsection{Model}

We suggest next to new minimal standard model type-II (NNMSM-II) by reducing the
 particle contents of the NNMSM\cite{hkt0}, which has two adjoint fermions 
$\lambda_a$ ($a=2,3$), four vector-like fermions $L_i'~(\overline{L_i'})$ 
($i=1,2$), the gauge singlet real scalar boson $S$, two right-handed neutrinos 
$N_i$, the inflaton $\varphi$, and the small cosmological constant $\Lambda$ in 
addition to the SM. Our model removes four vector-like fermions from the NNMSM 
 but adds new energy scale to the model. The quantum numbers of these particles 
are given in Table.\ref{tab1}. Only the gauge singlet scalar particle has 
odd-parity under an additional $Z_2$-symmetry while other additional particles 
have even-parity. We will show the singlet scalar becomes DM as in the NNMSM. 
Runnings of the gauge couplings are changed from the SM due to new particles 
with charges. The realization of the GCU is one of important results of this 
work as we will show later.

We consider the NNMSM-II as a renormalizable theory, and thus, the most general 
form of the Lagrangian allowed by the symmetries and renormalizability is given 
by
\begin{eqnarray}
 {\cal L}_{\rm NNMSM}   
  &=& {\cal L}_{\rm SM} + {\cal L}_S +{\cal L}_N
      +{\cal L}_\varphi+{\cal L}_\Lambda+{\cal L}', \\
 {\cal L}_{\rm SM} 
  &\supset& -\lambda\left(|H|^2-\frac{v^2}{2}\right)^2, \label{SM0} \\ 
 {\cal L}_S 
  &=& -\bar{m}_S^2S^2-\frac{k}{2}|H|^2S^2 - \frac{\lambda_S}{4!}S^4
      +(\text{kinetic term}), \label{S0} \\
 {\cal L}_N
  &=& -\left(\frac{M_{Ri}}{2}\overline{N_i^c}N_i
            +h_\nu^{i\alpha}\overline{N_i}L_\alpha\tilde{H}+c.c.\right)
     +(\text{kinetic term}), \label{neu} \\
 {\cal L_\varphi}
  &=& -B\varphi^4\left[\mbox{ln}\left(\frac{\varphi^2}{\sigma^2}\right)
                       -\frac{1}{2}\right]-\frac{B\sigma^4}{2}
      -\mu_1\varphi|H|^2-\mu_2\varphi S^2-\kappa_H\varphi^2|H|^2
      -\kappa_S\varphi^2S^2 \nonumber \\ 
  & & -(y_N^{ij}\varphi \overline{N_i}N_j
        +y_3\varphi\overline{\lambda_3}\lambda_3
        +y_2\varphi\overline{\lambda_2}\lambda_2+c.c.)
      +(\text{kinetic term}), \label{inf0} \\
 {\cal L}_\Lambda
  &=& (2.3\times10^{-3}\mbox{ eV})^4, \label{cc0} \\
 {\cal L}' &=& 
  \left(y_{\nu_\alpha}\overline{L_\alpha}\lambda_2\tilde{H}
  +M_3\overline{\lambda_3}\lambda_3+M_2\overline{\lambda_2}\lambda_2+h.c.\right)
  +(\text{kinetic terms}), \label{Yukawa0} 
\end{eqnarray}
with $\alpha=e,\mu,\tau$ and $\tilde{H}=i\sigma_2H^\ast$ where ${\cal L}_{SM}$ 
is the Lagrangian of the SM, which includes the Higgs potential. $v$ is the 
vacuum expectation value (VEV) of the Higgs as $v=246$ GeV. 
${\cal L}_{S,N,\varphi,\Lambda}$ are Lagrangians for the DM, right-handed 
neutrinos, inflaton, and the cosmological constant, respectively. 
${\cal L}_{\rm SM}+{\cal L}_{S,N,\Lambda}$ are the same as those of the 
NNMSM.\footnote{For the present cosmic acceleration, we simply assume that the 
origin of DE is the tiny cosmological constant, which is given in 
$\mathcal{L}_\Lambda$ of Eq.(\ref{cc0}), so that the NNMSM-II predicts the 
equation of state parameter as $\omega=-1$, like the NNMSM. We will not focus on
 the DE in this work anymore.} ${\cal L}'$ is new Lagrangian in the NNMSM-II 
where $L_\alpha$ is left-handed lepton doublets in the 
SM.\footnote{If one assigns odd-parity under $Z_2$ to $\lambda_2$ and does not 
introduce $S$, only the neutral component of $\lambda_2$ can be a DM candidate. 
The field content in the scenario without $S$ decreases compared to the NNMSM-II
 but it becomes difficult to satisfy the vacuum stability and triviality bounds 
with the center values of the Higgs and top masses at the same time. Since the 
$S$ also plays a crucial role for the discussions of the stability and 
triviality bounds in addition to the DM candidate as shown later, we focus on 
the model with $S$.}
\begin{table}[tbp]
\begin{center}
\begin{tabular}{c|ccccc} \hline \hline
           & $\lambda_3$ & $\lambda_2$ & $S$  & $N_i$ & $\varphi$ \\ \hline
 $SU(3)_C$ & 8           & 1           & 1    & 1     & 1         \\
 $SU(2)_L$ & 1           & 3           & 1    & 1     & 1         \\
\hline
 $Z_2$     & $+$         & $+$         & $-$  & $+$   & $+$       \\ 
\hline\hline
\end{tabular}
\end{center}
\caption{Quantum numbers of the additional particles in the NNMSM-II ($i=1,2$).}
\label{tab1}
\end{table}

\subsection{Gauge coupling unification}

At first, we investigate the runnings of the gauge couplings in the NNMSM-II. 
Since we introduce two adjoint fermions $\lambda_3$ and $\lambda_2$, listed in 
Tab.\ref{tab1}, the beta functions of the RGEs for the gauge couplings become  
 \begin{eqnarray}
  2\pi\frac{d\alpha_j^{-1}}{dt} &=& b_j, \label{g0}
 \end{eqnarray}
where $t\equiv\ln(\mu/1\mbox{ GeV})$, $\mu$ is the renormalization scale, and 
$\alpha_j\equiv g_j^2/(4\pi)$ $(j=1,2,3)$ with $g_1\equiv\sqrt{5/3}g'$. The beta
 functions from the SM and the new particles contribute to $b_j$. Each 
contribution from the SM and the new particles shown in Tab.\ref{tab1} is given
 by 
 \begin{eqnarray}
  && (b_1^{\rm SM},b_2^{\rm SM},b_3^{\rm SM})=(\frac{41}{10},-\frac{19}{6},-7),
     ~~~(b_1^{\lambda_3},b_2^{\lambda_3},b_3^{\lambda_3})=(0,0,2),~~~ 
        (b_1^{\lambda_2},b_2^{\lambda_2},b_3^{\lambda_2})=(0,\frac{4}{3},0).
 \end{eqnarray}
The NNMSM has assumed the same mass scale for the new particles as 
$M_{NP}=M_3=M_2=M_{L_i'}$ where $M_{L_i'}$ was the masses of the vector-like 
fermions. On the other hand, we allow different mass scales for the two adjoint 
fermions in the NNMSM-II. 

\begin{figure}
\begin{center}
\includegraphics[scale=1,bb=0 0 270 180]{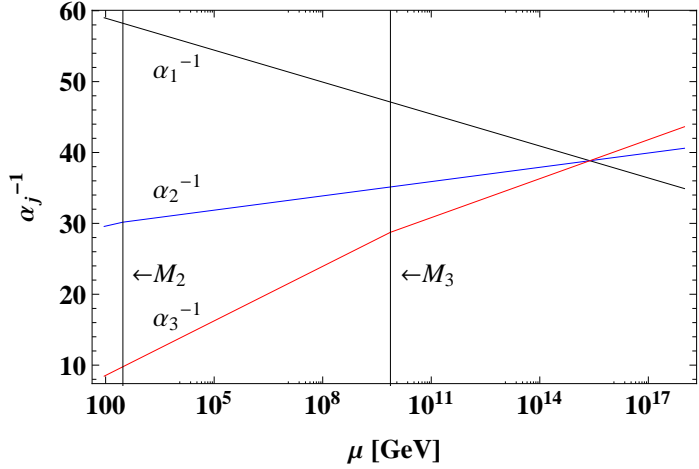}
\end{center}
\caption{The runnings of the gauge couplings in the NNMSM-II. The horizontal 
axis is the renormalization scale and the vertical axis is the values of 
$\alpha_i^{-1}$. The runnings of $\alpha_1^{-1}$, $\alpha_2^{-2}$, and 
$\alpha_3^{-1}$ are described by black, blue, and red solid curves, 
respectively. We take $M_3\simeq7.44\times10^9$ GeV and $M_2=300$ GeV, and the 
coupling unification is realized at $\Lambda_{\rm 
GCU}\simeq2.41\times10^{15}~{\rm GeV}$ GeV with $\alpha_{\rm GCU}^{-1}\simeq 
38.8$.}
\label{GCU-2-0}
\end{figure}

According to the numerical analyses, taking two masses as 
$M_3\simeq7.44\times10^9$ GeV and $M_2=300$ GeV can realize the GCU with a good 
precision at 1-loop level as shown in Fig.\ref{GCU-2-0}.\footnote{In this 
analysis, we take the following values as\cite{Beringer:1900zz}, 
$\sin^2\theta_W(M_Z)=0.231,~\alpha_{\rm em}^{-1}(M_Z)=128,$ and 
$\alpha_s(M_Z)=0.118,$ for the parameters in the EW theory, where $\theta_W$ is 
the Weinberg angle, $\alpha_{em}$ is the fine structure constant, and $\alpha_s$
 is the strong coupling, respectively.} 
The beta functions in the NNMSM-II are
 \begin{eqnarray}
  b_j=\left\{
       \begin{array}{ll}
        b_j^{\rm SM} & \mbox{ for }M_Z\leq\mu<M_2 \\
        b_j^{\rm SM}+b_j^{\lambda_2} & \mbox{ for }M_2\leq\mu<M_3 \\
        b_j^{\rm SM}+b_j^{\lambda_2}+b_j^{\lambda_3} & \mbox{ for }M_3\leq\mu
       \end{array}
      \right.,
 \end{eqnarray}
We show the thresholds of new particles with $M_3\simeq7.44\times10^9$ GeV 
and $M_2=300$ GeV by black solid lines. The realization of the GCU by adding 
these adjoint fermions was pointed out in Ref.\cite{Ibe:2009gt}. The NNMSM-II 
suggests the GCU at
 \begin{eqnarray}
  \Lambda_{\rm GCU}\simeq2.41\times10^{15}~{\rm GeV}, \label{LamGCU}
 \end{eqnarray}
with the unified coupling as 
 \begin{eqnarray}
  \alpha^{-1}\simeq38.8. \label{alGCU}
 \end{eqnarray}
A constraint from the proton decay experiments is 
$\tau(p\rightarrow\pi^0e^+)>8.2\times10^{33}$ years~\cite{Beringer:1900zz}. When 
we suppose the minimal $SU(5)$ GUT at $\Lambda_{\rm GCU}$, the protons decay of 
$p\rightarrow\pi^0e^+$ occurs by exchanging heavy gauge bosons of the GUT gauge 
group. The partial decay width of proton for $p\rightarrow\pi^0e^+$ is estimated
 as
 \begin{eqnarray}
  \Gamma(p\rightarrow\pi^0e^+)
   =\alpha_H^2\frac{m_p}{64\pi f_\pi^2}(1+D+F)^2
    \left(\frac{4\pi\alpha_{\rm GCU}}{\Lambda_{\rm GCU}}A_R\right)^2
    (1+(1+|V_{ud}|^2)^2),
 \end{eqnarray} 
where $\alpha_H^2$ is the hadronic matrix element, $m_p$ is the proton mass, 
$f_\pi$ is the pion decay constant, $D$ and $F$ are the chiral Lagrangian 
parameters, $A_R$ is the renormalization factor, and $V_{ud}$ is an element of 
the CKM matrix (e.g., see~\cite{Ibe:2009gt,Hisano:2000dg}). In our analysis, we 
take these parameters as $m_p=0.94$ GeV, $f_\pi=0.13$ GeV, $A_R\simeq1.02$, 
$D=0.80$, and $F=0.47$. A theoretical uncertainty on the proton life time comes 
from the hadron matrix elements as $\alpha_H=-0.0112\pm0.0034~{\rm 
GeV}^3$\cite{Aoki:2008ku}. When $\alpha_H$ is taken as a smaller value, which is
 $\alpha_H=-0.0146~{\rm GeV}^3$, the proton life time is predicted as small, and
 this case gives a conservative limit. At the point determined by (\ref{LamGCU})
 and (\ref{alGCU}), the proton life time can be evaluated as 
$\tau\simeq5.19\times10^{33}~(1.82\times10^{34})$ years for 
$\alpha_H=-0.0146~(-0.0078)~{\rm GeV}^3$. Thus, the value of 
$\alpha_H=-0.0078~{\rm GeV}^3$ can satisfy the experimental bound from the 
proton decay although the conservative case ($\alpha_H=-0.0146~{\rm GeV}^3$) can
not do it. For the center value of $\alpha_H=-0.0112~{\rm GeV}^3$, the proton 
life time is evaluated as $\tau\simeq8.55\times10^{33}$ years, which can also 
satisfy the experimental limit. Since the future Hyper-Kamiokande experiment is 
expected to exceed the life time $\mathcal{O}(10^{35})$ years\cite{Abe:2011ts}, 
which corresponds to $\Lambda_{\rm GCU}\simeq4.42_{-0.73}^{+0.63}\times10^{15}$ GeV 
for $\alpha_H=-0.0112\pm 0.0034~{\rm GeV}^3$, the proton decay is observed if 
the NNMSM-II is true.

When we take a larger value of $M_2$ such as 800 GeV, the GCU can be realized 
with a larger value of $M_3$ as $2.03\times10^{10}$ GeV. However, the GCU with 
larger value of $M_2$ leads to smaller scale of $\Lambda_{\rm GCU}$, and thus, the
 constraint from the proton decay becomes stronger. In fact, the GCU with 
$M_2=800$ GeV leads to the proton life time as 
$\tau\simeq(2.14-7.48)\times10^{33}$ years for $\alpha_H=-0.0112\pm 0.0034~{\rm
 GeV}^3$, which is ruled out by the proton decay experiment. Therefore, the 
upper bounds on the adjoint fermions masses are 
$(M_3,M_2)\lesssim(2\times10^{10},800)$ GeV. On the other hand, the LHC 
experiment gives a lower limit to the $SU(2)_L$ triplet mass $M_2$ as 245 
GeV$\leq M_2$ by the ATLAS examining the channel of four lepton final states 
(and $(180-210)~{\rm GeV}\leq M_2$ by the CMS for the three lepton final 
states)\cite{CMS:2012ra}. As a result, the allowed region for $M_2$ in the 
NNMSM-II is 245 GeV$\leq M_2\lesssim800$ GeV. The corresponding region for $M_3$
 is $7\times10^9~{\rm GeV}\lesssim M_3\lesssim2\times10^{10}$ GeV in order to 
realize the GCU. Since the $SU(2)_L$ triplet fermion can be discovered up to 
$M_2\leq750$ GeV by the LHC with $\sqrt{s}=14$ TeV~\cite{delAguila:2008cj}, most
 of the mass region of $M_2$ in the NNMSM-II can be checked by the experiment.

\subsection{Abundance and stability of new fermions}

Next, we discuss an abundance and stability of new fermions, $\lambda_3$ and 
$\lambda_2$. $\lambda_3$ is expected to be long lived since it cannot decay into
 the SM sector. A stable colored particle is severely 
constrained by experiments with heavy isotopes, since it bounds in nuclei and 
appears as anomalously heavy isotopes (e.g., see~\cite{Giudice:2004tc}). The 
number of the stable colored particles per nucleon should be smaller than 
$10^{-28}~(10^{-20})$ for its mass up to 1 (10) 
TeV\cite{Smith:1982qu,Hemmick:1989ns}. But the calculation of the relic 
abundance of the stable colored particle is uncertain because of the dependence 
on the mechanism of hadronization and nuclear binding\cite{Baer:1998pg}.

In this paper, we apply the same scenario in order to avoid the problem of the 
presence of the stable colored particle as in the NNMSM. It is to consider few 
production scenario for the stable particle, i.e., if the stable particle were 
rarely produced in the thermal history of the universe and clear the constraints
 of the colored particle. In fact, a particle with mass of $M$ is very rarely 
produced thermally if the reheating temperature after the inflation is lower 
than $M/(35\sim40)$.\footnote{We thank S. Matsumoto for pointing out this point 
in a private discussion.} Therefore, we consider a reheating temperature 
$T_{RH}$ as,
 \begin{eqnarray}
  T_{RH}\lesssim\frac{M_3}{40}=1.86\times10^8\mbox{ GeV},
 \end{eqnarray}
since $M_3=7.44\times10^9$ GeV. On the other hand, $\lambda_2$ is thermally 
produced but it can decay into the SM particles through the Yukawa interaction 
in Eq.(\ref{Yukawa0}) before the Big Bang nucleosynthesis (BBN). The condition 
for the decay before the BBN is $\mathcal{O}(10^{-13})\lesssim|y_{\nu_\alpha}|$, 
which can also be consistent with a constraint from a neutrino mass generation 
as we will discuss later. Therefore, the presence of two new adjoint fermions in
 the NNMSM-II for the GCU is not problematic.

\subsection{Stability, triviality, and dark matter}

In this section, we investigate parameter region where not only stability and 
triviality bounds but also correct abundance of the DM are achieved. The 
ingredients of Higgs and DM sector in the NNMSM-II are the same as the 
NNMSM\cite{hkt0}, which are given by $\mathcal{L}_{\rm SM}$ and $\mathcal{L}_S$ in
 Eqs.(\ref{SM0}) and (\ref{S0}) but the runnings of the gauge couplings are 
different with those of NNMSM. The singlet scalar $S$ becomes the DM also in the
 NNMSM-II. Ref.\cite{hkt0} has pointed out that there are two typical regions in
 which the vacuum stability, the correct abundance of DM, 126 GeV Higgs mass, 
and the latest experimental value of the top mass can be realized at the same 
time. One is the lighter mass region of DM and the other is heavier one. It has 
been also shown the top mass dependence is quite strong even within the 
experimental error of top pole mass, $M_t=173.5\pm1.4$ GeV. The future XENON100 
experiment with 20 times sensitivity will be able to rule out the lighter mass 
region completely. On the other hand, the heavier mass region can be currently 
allowed by all experiments searching for DM. The region will be ruled out or 
checked by the future direct experiments of XENON100$\times$20, XENON1T and/or 
combined data from indirect detections of Fermi$+$CTA$+$Planck at $1\sigma$ CL.

In the NNMSM-II, the running of gauge couplings are slightly changed with those 
of the NNMSM. Therefore, we reanalyze the vacuum stability, triviality, and the 
correct abundance of DM in the NNMSM-II. The RGEs for three quartic couplings of
 the scalars\cite{Davoudiasl:2004be} and the DM mass are given by,
 \begin{eqnarray}
  (4\pi)^2\frac{d\lambda}{dt} 
   &=& 24\lambda^2+12\lambda y^2-6y^4-3\lambda(g'{}^2+3g^2)
       +\frac{3}{8}\left[2g^4+(g'{}^2+g^2)^2\right]+\frac{k^2}{2}, \label{lam0} 
\\
  (4\pi)^2\frac{dk}{dt} 
   &=& k\left[4k+12\lambda+\lambda_S+6y^2-\frac{3}{2}(g'{}^2+3g^2)\right], 
       \label{k0} 
\\
  (4\pi)^2\frac{d\lambda_S}{dt} &=& 3\lambda_S^2+12k^2. \label{h0}
 \end{eqnarray}
and
 \begin{eqnarray}
  m_S=\sqrt{\bar{m}_S^2+kv^2/8},
 \end{eqnarray}
respectively. We should also use the present limits for the singlet DM model. We
 comment on Eq.(\ref{k0}) that right-hand side of the equation is proportional 
to $k$ itself. Thus, if we take a small value of $k(M_Z)$, evolution of $k$ 
tends to be slow and remained in a small value, and the running of $\lambda$ 
closes to that of SM.
In our 
analysis, boundary conditions of the Higgs self-coupling and top Yukawa coupling
 are given by
\begin{eqnarray}
  \lambda(M_Z)=\frac{m_h^2}{2v^2}=0.131,~~~
  y(M_t)=\frac{\sqrt{2}m_t(M_t)}{v} \label{BC}
 \end{eqnarray}
for the RGEs, where the VEV of the Higgs field is $v=246~{\rm GeV}$.

\begin{figure}
\subfigure[]{
\includegraphics[clip,width=0.465\columnwidth]{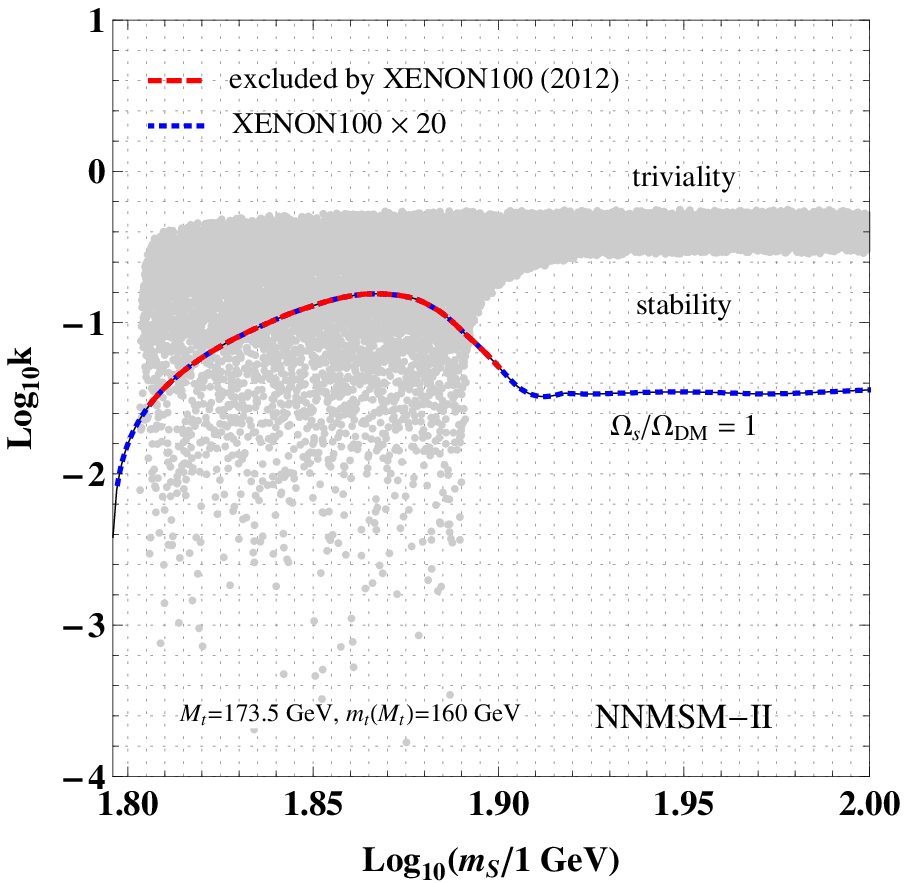}
}
\subfigure[]{
\includegraphics[clip,width=0.45\columnwidth]{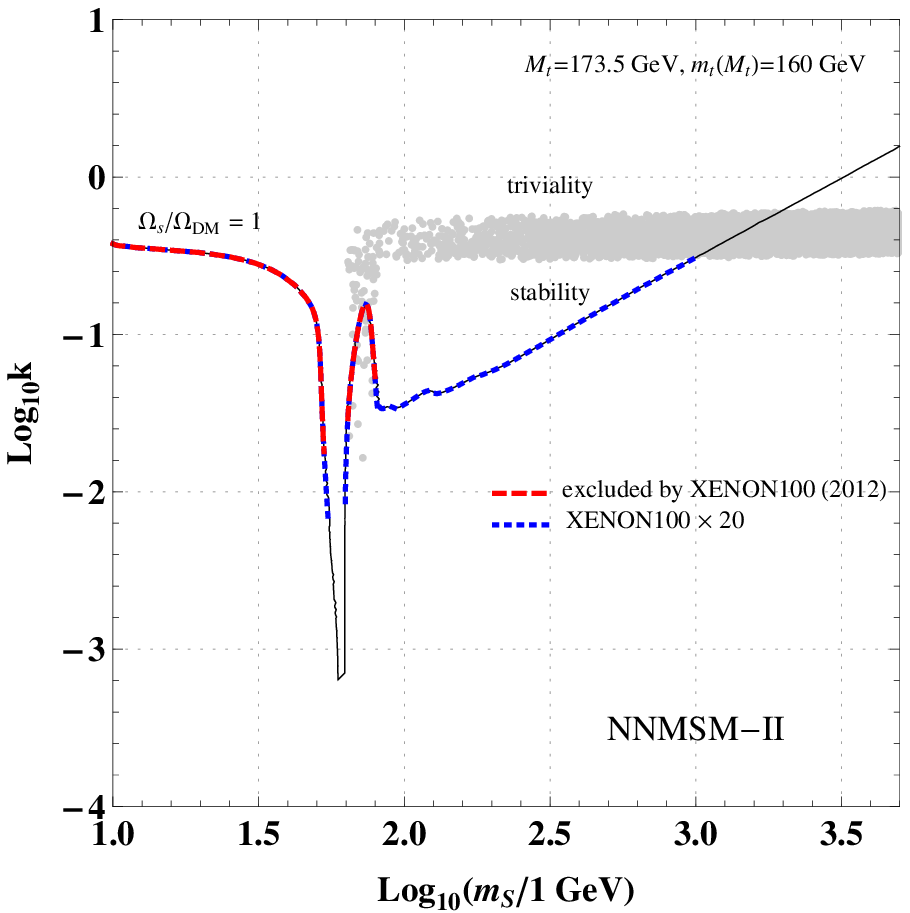}
}
\caption{
A contour of fixed relic density $\Omega_S/\Omega_{\rm DM}=1$ and a 
region, which is described by gray plots, satisfying the stability and 
triviality bounds with $M_t=173.5$ GeV ($m_t(M_t)=160$ GeV) in the NNMSM-II. The
 boundaries of the plotted region are determined by the stability and triviality
 conditions. The words ``stability" and ``triviality" in both figures mean the 
corresponding conditions. The (red) dashed and (blue) dotted lines are 
experimental limits from XENON100 (2012) and 20 times sensitivity of XENON100, 
respectively. (a) The mass region is $63~{\rm GeV}\leq m_S\leq100~{\rm GeV}$ 
($1.8\leq\log(m_S/1~{\rm GeV})\leq2.0$). (b) The mass region is $10~{\rm 
GeV}\leq m_S\leq 5000~{\rm GeV}$ ($1.0\leq\log(m_S/1~{\rm GeV})\leq3.7$). 
}
\label{omega_type2_2}
\end{figure}

Let us solve the RGEs, Eqs.(\ref{lam0})$\sim$(\ref{h0}), and obtain the stable 
solutions, i.e., the scalar quartic couplings are within the range of 
$0<(\lambda,k,\lambda_S)<4\pi$ up to the Planck scale $M_{\rm pl}=10^{18}$ GeV. 
Figure \ref{omega_type2_2} shows the case of central value as $M_t=173.5~{\rm 
GeV}$ (leading to $m_t(M_t)=160~{\rm GeV}$). 
The solutions of the RGEs are described by gray plots in 
Fig.\ref{omega_type2_2}, where the horizontal and vertical axes are 
$\log_{10}(m_S/1\mbox{ GeV})$ and $\log_{10}k$ at the $M_Z$ scale, respectively. 
The boundaries of the plotted region are determined by the stability and 
triviality conditions. The words ``stability" and ``triviality" in both figures 
mean the corresponding conditions. We also show the contour satisfying 
$\Omega_S/\Omega_{\rm DM}=1$ with $\Omega_{\rm DM}=0.115$, where $\Omega_S$ and 
$\Omega_{\rm DM}$ are density parameter of the singlet DM and observed value of 
the parameter\cite{Bennett:2012zja}, respectively. The contour is calculated by 
{\tt micrOMEGAs}\cite{Belanger:2013oya}. Since there is no DM candidate except 
for the $S$ to compensate $\Omega_S/\Omega_{\rm DM}<1$, which is above the 
contour, we focus only on the contour. The relic density depends on $k$ and 
$m_S$ but not $\lambda_S$, meanwhile $\lambda_S$ affects the stability and 
triviality bounds. In the figure, $\lambda_S(M_Z)$ is randomly varied from 0 to 
$4\pi$, where $\lambda_S$-dependence of the stability and triviality bounds is 
not stringent, and most of $\lambda_S(M_Z) \in [0,1]$ as the boundary condition 
can satisfy the bounds. A direct DM search experiment, XENON100 (2012), gives 
an exclusion limit\cite{Cline:2013gha}, which is described by the (red) dashed 
line in Fig.\ref{omega_type2_2}.\footnote{The lighter DM mass region as 
$m_S\lesssim62.5$ GeV ($\log(m_S/1~{\rm GeV})\lesssim1.8$) with 
$-1.8\lesssim\log_{10}k$ is ruled out by the invisible Higgs decay into a pair of
 DM at the LHC~\cite{Cline:2013gha}. But the stability and triviality can not be
 realized in the region as shown in Fig.\ref{omega_type2_2}.} There are two 
regions, $R_{1,2}$, which satisfy both the correct DM abundance and the 
triviality bound simultaneously,
\begin{eqnarray}
 R_1
&=&\left\{
\begin{array}[tb]{l l}
63.5~{\rm GeV}\lesssim m_S \lesssim 64.0~{\rm GeV} 
&
 (1.803\lesssim\log_{10}(m_S/1~{\rm GeV})\lesssim 1.806)
\\
2.40\times10^{-2}\lesssim k(M_Z) \lesssim 2.63\times10^{-2} 
&
 (-1.64\lesssim\log_{10}k(M_Z)\lesssim -1.58)
\end{array}
\right.,\\
R_2
&=&
\left\{
\begin{array}[tb]{l l}
955~{\rm GeV}\lesssim m_S \lesssim 2040~{\rm GeV} & (2.98\lesssim\log_{10}(m_S/1~{\rm GeV})\lesssim 3.31) \\
0.316\lesssim k(M_Z)\lesssim0.631 &
 (-0.50\lesssim\log_{10}k(M_Z)\lesssim-0.20)
\end{array}
\right..
\end{eqnarray}
The future XENON100 experiment with 20 times sensitivity, which is described by 
the (blue) dotted lines in Fig.\ref{omega_type2_2}, will be able to rule out the
 lighter $m_S$ region $R_1$, completely. On the other hand, the heavier $m_S$ 
region, $R_2$, can be currently allowed by all experiments searching for DM. It 
is seen that the future XENON100$\times$20 can check up to $m_S\lesssim$ 1000 
GeV ($\log_{10}(m_S/1\mbox{ GeV})\lesssim3$). The future XENON1T experiment and 
combined data from indirect detections of Fermi$+$CTA$+$Planck at $1\sigma$ CL 
may be able to reach up to $m_S\simeq5$ TeV\cite{Cline:2013gha}. The lower and 
upper bound on $m_S$ in the region $R_1$ come from the triviality bound on 
$\lambda$ and XENON100 (2012) experiment, respectively. On the other hand, the 
lower and upper bound on $m_S$ in the region $R_2$ are given by the stability 
and triviality bounds on $\lambda$, respectively. Since $k$ in the R.H.S of 
Eq.(\ref{lam0}) is effective only above the energy scale of $m_S$, the 
triviality bound on $\lambda$ becomes severe as the $m_S$ becomes small. We can 
also find that both the regions $R_1$ and $R_2$ are almost the same as in the 
NNMSM\cite{hkt0}. This means that the differences of runnings of the gauge 
couplings do not affect on the stability, triviality, and correct abundance of 
DM in these class of model. The favored region still depends on the top mass 
rather than the runnings of the gauge couplings as pointed out in 
Ref.\cite{hkt0}. In fact, the heavier mass region becomes narrow as the top mass
 becomes larger due to the stability bound while the region $R_1$ does not 
depends on the top mass because the triviality bound on $\lambda$ does not 
depend on the top Yukawa coupling (see Ref.\cite{hkt0} for the detailed 
discussions about the top mass dependence of the region).

\subsection{Inflation, neutrinos, and baryogenesis}

In this section, realizations of the inflation, suitable tiny active neutrino 
mass, and baryogenesis are discussed. The relevant Lagrangian for the inflaton 
is given by $\mathcal{L}_\varphi$ in Eq.(\ref{inf0}). The 
WMAP~\cite{Bennett:2012zja,Hinshaw:2012aka} and the Planck~\cite{Ade:2013zuv} 
measurements of the cosmic microwave background (CMB) constrain the cosmological
 parameters related with the inflation in the early universe. In particular, the
 first results based on the Planck measurement with a WMAP polarization 
low-multipole likelihood at $\ell\leq23$ 
(WP)~\cite{Bennett:2012zja,Hinshaw:2012aka} and high-resolution (highL) CMB data
 gives
 \begin{eqnarray}
  n_s &=& 0.959\pm0.007~(68\%;~\mbox{Planck$+$WP$+$highL}), \\
  r_{0.002} &<& \left\{ 
                 \begin{array}{ll}
                  0.11 & (95\%;~\mbox{no running},~\mbox{Planck$+$WP$+$highL}) \\
                  0.26 & (95\%;~\mbox{including running},~\mbox{Planck$+$WP$+$highL})
                 \end{array}
                \right., \\
  dn_s/d\mbox{ln}k &=& -0.015\pm0.017~(95\%;~\mbox{Planck$+$WP$+$highL}),
 \end{eqnarray}
for the scalar spectrum power-law index, the ratio of tensor primordial power to
 curvature power, the running of the spectral index, respectively, in the 
context of the $\Lambda$CDM model. Regarding $r_{0.002}$, the constraints are 
given for both no running and including running cases of the spectral indices.

We also adopt the same inflation model in the NNMSM-II as in the NNMSM. The 
inflaton potential is the Coleman-Weinberg (CW) 
type\cite{Coleman:1973jx,Knox:1992iy}. In this potential Eq.(\ref{inf0}), the 
VEV of $\varphi$ becomes $\sigma$. When we take 
$(\phi,\sigma,B)\simeq(6.60\times10^{19}\mbox{ 
GeV},9.57\times10^{19}$ GeV$,10^{-15})$, the model can lead to $n_s=0.96$, 
$r=0.1$, $dn_s/d\mbox{ln}k\simeq8.19\times10^{-4}$, and 
$(\delta\rho/\rho)\sim\mathcal{O}(10^{-5})$, which are consistent with the 
cosmological data. The values of couplings of inflaton with the Higgs, DM, 
right-handed neutrinos, and new adjoint fermions are also constrained because 
there is an upper bound on the reheating temperature after the inflation as 
$T_{RH}\lesssim1.86\times10^8\mbox{ GeV}$. This upper bound leads to 
$\mu_{1,2}\lesssim6.86\times10^7$ GeV and 
$(y_N^{ij},y_3,y_2)\lesssim1.79\times10^{-6}$. Since $\kappa_{H,S}$ should be 
almost vanishing at the low energy for the realizations of the EW symmetry 
breaking and the DM mass, we take the values of $\kappa_{H,S}$ as very tiny at 
the epoch of inflation. The smallness of $\kappa_{H,S}$ does not also spoil the 
stability and triviality bounds. As for the lower bound of the reheating 
temperature, it depends on the baryogenesis mechanism. When the baryogenesis 
works through the sphaleron process, the reheating temperature must be at least 
higher than $\mathcal{O}(10^2)$ GeV. 

The neutrino sector is shown in Eq.(\ref{neu}), where tiny active neutrino 
masses are obtained through the type-I and III seesaw mechanisms. Since there 
are two right-handed neutrinos and one adjoint ($SU(2)_L$ triplet) fermion, 
three active neutrinos are predicted to be massive in the NNMSM-II. The Yukawa 
coupling of the triplet fermion should be 
$|y_{\nu_\alpha}|\lesssim\mathcal{O}(10^{-6})$ not to exceed a typical neutrino mass
 scale as $m_\nu\sim0.1$ eV. Thus, the region of 
$\mathcal{O}(10^{-13})\lesssim|y_{\nu_\alpha}|\lesssim\mathcal{O}(10^{-6})$ is allowed
 in which the lower bound comes from the discussion of the BBN as mentioned 
above. Reminding the reheating temperature in the NNMSM-II, masses of the 
right-handed neutrino must be lighter than $1.86\times10^8$ GeV. What mechanism 
can induce the suitable baryon asymmetry in such a low reheating temperature? 
One possibility is the resonant leptogenesis\cite{Pilaftsis:2003gt} in which the
 right-handed neutrinos can be light up to $1$ TeV. Thus, the reheating 
temperature, 1 TeV$\lesssim T_{RH}\lesssim1.86\times10^8$ GeV, can realize the 
resonant leptogenesis, which means the couplings of inflaton as $369\mbox{ 
GeV}\lesssim\mu_{1,2}\lesssim6.86\times10^7$ GeV and 
$9.63\times10^{-12}\lesssim(y_N^{ij},y_3,y_2)\lesssim1.79\times10^{-6}$ in 
Eq.(\ref{inf0}). 

\section{NNMSM type-III}

\subsection{Models}

We also discuss other possibilities, which are alternatives to NNMSM and 
NNMSM-II, for realizing the GCU by different particle contents with them. Here, 
we suggest a class of model with several generations of the adjoint fermions and
 without the right-handed neutrinos. We refer this class of models to the 
NNMSM-III. We focus on two simple models, NNMSM-III-A and B, in this class of 
models. Both the NNMSM-III-A and B introduces three new fields such as one 
$\lambda_3$ and two generations of $\lambda_{2,i}$ ($i=1,2$) in addition to the 
singlet DM, inflaton, and the cosmological constant. Then, the NNMSM-III-A 
requires the mass scales of three new fields are the same and the NNMSM-III-B 
allows different mass scales between $\lambda_3$ and $\lambda_{2,i}$. The 
quantum numbers of these particles for both models are given in 
Table.~\ref{tab2}.
\begin{table}[tbp]
\begin{center}
\begin{tabular}{c|cccc} \hline \hline
           & $\lambda_3$ & $\lambda_{2,i}$ & $S$ & $\varphi$ \\ \hline
 $SU(3)_C$ & 8           & 1               & 1     & 1         \\
 $SU(2)_L$ & 1           & 3               & 1     & 1         \\ \hline
 $Z_2$     & $+$         & $+$         & $-$     & $+$       \\ 
\hline\hline
\end{tabular}
\end{center}
\caption{Quantum numbers of additional particles in the NNMSM-III-A and B 
($i=1,2$).}
\label{tab2}
\end{table}
Only the singlet scalar DM has odd-parity under an additional $Z_2$ symmetry 
like the NNMSM-II. Since the runnings of the GCU are still changed from the 
previous models, we will focus on them later.

The NNMSM-III-A and B are also presented as renormalizable theory. The most 
general form of the Lagrangian allowed by the symmetries and renormalizability 
is given by
\begin{eqnarray}
 {\cal L}_{\rm NNMSM}   
  &=& {\cal L}_{\rm SM} + {\cal L}_S+{\cal L}_\varphi+{\cal L}_\Lambda+{\cal L}', \\
 {\cal L_\varphi}
  &=& -B\varphi^4\left[\mbox{ln}\left(\frac{\varphi^2}{\sigma^2}\right)
                       -\frac{1}{2}\right]-\frac{B\sigma^4}{2}
      -\mu_1\varphi|H|^2
      -\mu_2\varphi S^2-\kappa_H\varphi^2|H|^2
      -\kappa_S\varphi^2S^2 \nonumber \\ 
  & & -(y_3\varphi\overline{\lambda_3}\lambda_3
        +y_2^{ij}\varphi\overline{\lambda_{2i}}\lambda_{2j}+c.c.)
      +(\text{kinetic term}), \label{inf} \\
 {\cal L}' &=& 
  -\left(y_{\nu_\alpha}^{i}\overline{L_\alpha}\lambda_{2,i}\tilde{H}
  +M_3\overline{\lambda_3}\lambda_3+M_{2i}\overline{\lambda_{2i}}\lambda_{2i}
  +h.c.\right)+(\text{kinetic terms}), \label{Yukawa} 
\end{eqnarray}
with
 \begin{eqnarray}
  \begin{array}{ll}
   M_{NP}\equiv M_3=M_{2,i} & \mbox{for NNMSM-III-A} \\
   M_3\neq M_{2,i},~~M_{2,1}=M_{2,2} & \mbox{for NNMSM-III-B}
  \end{array},
 \end{eqnarray}
where the inflaton potential in ${\cal L}_\varphi$ is the same as those of the 
other NNMSMs but the interactions are different with them. Then, ${\cal L}'$ is 
new Lagrangian for the models. The mass matrix $M_2$ is assumed to be diagonal, 
for simplicity. The Lagrangians ${\cal L}_{\rm SM}+{\cal L}_S+{\cal L}_\Lambda$ 
are the same as those of the NNMSMs.\footnote{We also simply assume that the 
origin of DE is the tiny cosmological constant as in the other NNMSMs.} One of 
advantages of the NNMSM-III-A is that we does not need two right-handed 
neutrinos and different mass scales between $\lambda_3$ and $\lambda_{2,i}$ for 
the realization of the GCU, unlike the NNMSM-II as we will show later. Thus, we 
introduce one mass scale for the new adjoint fermions and define it as 
$M_{NP}\equiv M_3=M_{2,i}$ for NNMSM-III-A. The NNMSM-III-B is a generalization 
of NNMSM-III-A.

\subsection{Gauge coupling unification}

At first, we investigate the runnings of the gauge couplings in the NNMSM-III-A 
and B. Since we introduce three adjoint fermions $\lambda_3$ and $\lambda_{2,i}$
 $(i=1,2)$ listed in Tab.~\ref{tab2}, the beta functions for the gauge couplings 
are modified to 
\begin{eqnarray}
  && (b_1^{\rm SM},b_2^{\rm SM},b_3^{\rm SM})=(\frac{41}{10},-\frac{19}{6},-7),
     ~~(b_1^{\lambda_3},b_2^{\lambda_3},b_3^{\lambda_3})=(0,0,2),~~
        (b_1^{\lambda_2},b_2^{\lambda_2},b_3^{\lambda_2})=(0,\frac{8}{3},0).
  \label{b-III}
 \end{eqnarray}
\subsubsection{NNMSM-III-A case}
\begin{figure}
\begin{center}
\includegraphics[scale=1,bb=0 0 270 180]{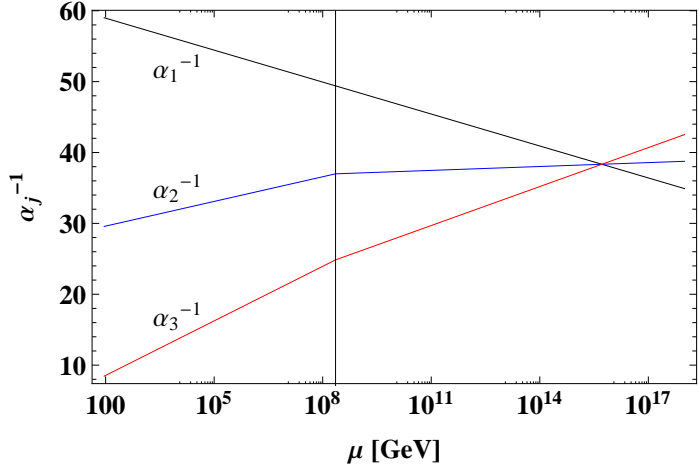}
\end{center}
\caption{The runnings of the gauge couplings in the NNMSM-III-A. The meanings of
 the figure is the same as in the Fig.\ref{GCU-2-0}. We take $M_{\rm 
NP}=2.26\times 10^8$ GeV, and the coupling unification is realized at 
$\mu=\Lambda_{\rm GUT}\simeq5.20\times 10^{15}$ GeV with $\alpha_{\rm 
GCU}^{-1}\simeq 38.3$.}
\label{fig2}
\end{figure}

According to the numerical analyses, taking a free parameter $M_{NP}$ as 
$2.26\times10^8$ GeV in the NNMSM-III-A can realize the GCU with a good 
precision at 1-loop level as shown in Fig.\ref{fig2}.\footnote{We take the same
 values of parameters in the EW theory as in the previous model.}
Since all masses of new adjoint fermions are around the same scale, 
$M_Z<M_{\rm NP}=M_3=M_2$, we should utilize the RGEs of Eq.(\ref{g0}) with 
$b_j^{\rm SM}+b_j^{\lambda_3}+b_j^{\lambda_2}$ given in Eq.(\ref{b-III}) at high
 energy scale ($M_{NP}\leq\mu$) while the right-handed side of Eq.(\ref{g0}) 
must be $b_j^{\rm SM}$  at low energy scale ($M_Z\leq\mu<M_{NP}$). We show the 
threshold of new particles with $2.26\times10^8$ GeV mass by a black solid line.
 The NNMSM-III-A suggests the GCU at
 \begin{eqnarray}
  \Lambda_{\rm GCU}\simeq5.20\times10^{15}~{\rm GeV}
 \end{eqnarray}
with the unified coupling as
 \begin{eqnarray}
  \alpha_{\rm GCU}^{-1}\simeq38.3.
 \end{eqnarray}
The model also predicts the proton life-time as 
$\tau=1.94_{-0.80}^{+2.06}\times10^{35}$ years for 
$\alpha_H=-0.0112\pm0.0034~{\rm GeV}^3$. Therefore, this model can satisfy the 
constraint on the proton decay even with the most conservative value of 
$\alpha_H$ but it might be difficult to check the proton decay in this model by 
the future Hyper-Kamiokande experiment.

There are other initial setups for realizing the GCU with the same mass scales 
for new adjoint fermions, i.e., more generations of adjoint fermions can also 
lead to the GCU. Some examples are given in Appendix.

\subsubsection{NNMSM-III-B case}
\begin{figure}
\begin{center}
\includegraphics[scale=1,bb=0 0 270 180]{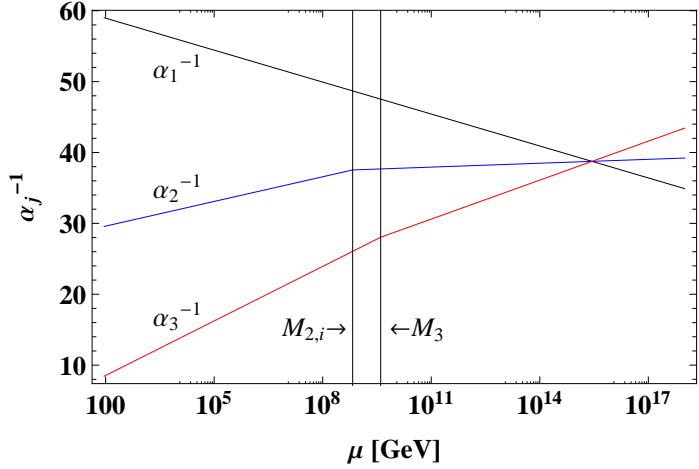}
\end{center}
\caption{The runnings of the gauge couplings in the NNMSM-III-B. The meanings of
 the figure is the same as in the Figs.\ref{GCU-2-0} and \ref{fig2}. We take 
$M_3\simeq4\times10^9$ GeV and $M_{2,i}\simeq6.73\times10^8$ GeV, and the 
coupling unification is realized at 
$\mu=\Lambda_{\rm GUT}\simeq2.77\times 10^{15}$ GeV with $\alpha_{\rm 
GCU}^{-1}\simeq 38.8$.}
\label{fig3}
\end{figure}

Since one can generally take different mass scales between $M_3$ and $M_{2,i}$, 
we consider the simplest case in the generalization, which is the NNMSM-III-B. 
According to the numerical analyses, when we take $M_3\simeq4\times10^9$ GeV and
 $M_{2,i}\simeq6.73\times10^8$ GeV, the GCU can be realized as shown in 
Fig.~\ref{fig3}.
The beta functions become in the NNMSM-II are
 \begin{eqnarray}
  b_j=\left\{
       \begin{array}{ll}
        b_j^{\rm SM} & \mbox{ for }M_Z\leq\mu<M_{2,i} \\
        b_j^{\rm SM}+b_j^{\lambda_2} & \mbox{ for }M_{2,i}\leq\mu<M_3 \\
        b_j^{\rm SM}+b_j^{\lambda_2}+b_j^{\lambda_3} & \mbox{ for }M_3\leq\mu
       \end{array}
      \right.,
 \end{eqnarray}
with Eq.(\ref{b-III}). We show the thresholds of new particles with 
$M_3\simeq4\times10^9$ GeV and 
$M_{2,i}\simeq6.73\times10^8$ GeV masses by black solid lines. This case 
suggests the GCU at 
 \begin{eqnarray}
  \Lambda_{\rm GCU}\simeq2.77\times10^{15}~{\rm GeV}
 \end{eqnarray}
with the unified coupling as 
 \begin{eqnarray}
  \alpha_{\rm GCU}^{-1}\simeq38.8.
 \end{eqnarray}
This mass spectrum interestingly predicts the proton life time as 
$\tau=1.50_{-0.62}^{+1.60}\times10^{34}$ years for 
$\alpha_H=-0.0112\pm0.0034~{\rm GeV}^3$. Therefore, the model with this mass 
spectrum can satisfy the constraint on the proton decay even with the most 
conservative value of $\alpha_H$ and it can be checked by the future 
Hyper-Kamiokande experiment if the model is true. This is one advantage 
compared to the NNMSM-III-A. In fact, there is an upper bound on the mass scale 
of $M_3$ (or $M_{2,i}$), which just comes from the proton decay. Therefore, in a
 broad region of $M_3\lesssim4\times10^9$ GeV (or 
$M_{2,1}\lesssim6.73\times10^8$ GeV), there are solutions for the realization of
 the GCU. There are also initial setups for the GCU with different mass scales 
between $\lambda_3$ and $\lambda_{2,i}$ and more generations of them. Some 
simple examples are given in Appendix.

\subsection{Abundance and stability of new fermions}

We also adopt the few production scenario for the colored particle $\lambda_3$ 
in the NNMSM-III in order to avoid the problem of the presence of the colored 
particle. The reheating temperature should be 
 \begin{eqnarray}
  T_{RH}<\left\{
          \begin{array}{ll}
           5.65\times10^6~{\rm GeV} & \mbox{ for the NNMSM-III-A} \\
           10^8~{\rm GeV} & \mbox{ for the NNMSM-III-B}
          \end{array}
         \right..
 \end{eqnarray}
Note that $\lambda_{2,i}$ are not also thermally produced because of 
$M_{\lambda_{2,i}}>T_{RH}$ in the models.

\subsection{Stability, triviality, and dark matter}

Regarding the stability, triviality, and DM, since the 
differences of the runnings of the gauge couplings do not almost affect the 
stability and triviality bounds, the favored regions in the NNMSM-III-A and B 
are also almost the same as in the NNMSM-II. Thus, the NNMSM-III-A and B predict
 the same mass region of DM and $k$ as in the other NNMSMs.

\subsection{Inflation, neutrinos, and baryogenesis}

Realizations of the inflation, suitable tiny active neutrino mass and 
baryogenesis in the NNMSM-III are discussed in this section. Regarding the 
inflation model, the same CW type inflaton potential as in the previous models 
is utilized in the models but the upper bounds on the inflaton couplings are 
changed due to different constraint on the reheating temperature. The upper 
bounds on the inflaton couplings are 
 \begin{eqnarray}
  \mu_{1,2} &\lesssim& \left\{
                        \begin{array}{ll}
                         2.09\times10^6~{\rm GeV} & \mbox{ for NNMSM-III-A} \\
                         3.69\times10^7~{\rm GeV} & \mbox{ for NNMSM-III-B}
                        \end{array}
                       \right. \\
  y_3,y_2
   &\lesssim& \left\{
               \begin{array}{ll}
                5.44\times10^{-8} & \mbox{ for NNMSM-III-A} \\
                9.63\times10^{-7} & \mbox{ for NNMSM-III-B}
               \end{array}
              \right..
 \end{eqnarray}

Since the NNMSM-III does not include the right-handed neutrinos, the neutrino 
sector is changed from the NNMSM-II. In the NNMSM-III, the tiny active neutrino 
mass can be realized by $SU(2)_L$ adjoint fermions through the type-III seesaw 
mechanism. The relevant Lagrangian is given by Eq.(\ref{Yukawa}). Since both 
NNMSM-III-A and B have only two generations of $\lambda_{2,i}$, one of active 
neutrinos is predicted to be massless $m_1=0$ ($m_3=0$) for the normal 
(inverted) mass hierarchy. Reminding the condition $M_{\lambda_{2,i}}>T_{RH}$ in the 
NNMSM-III-A and B for the few production scenario, $\lambda_{2,i}$ do not play a 
role for generating the baryon asymmetry of the universe. What mechanism can 
induce the BAU? One possibility is the baryogenesis from the dark 
sector\cite{Haba:2010bm} with supposing an asymmetry between DM and anti-DM in 
which the asymmetry of the dark matter sector including a new dark matter number
 can be converted into the lepton number.\footnote{This mechanism can be 
employed in the other NNMSMs if one supposes the above asymmetry for the DM 
sector in the models.} As a result, the baryon number can be generated through 
the sphaleron process. This means that the reheating temperature should be 
typically $\mathcal{O}(10^2)~{\rm GeV}\lesssim T_{RH}$ for both NNMSM-III-A and 
B, which leads to $\mathcal{O}(10)~{\rm GeV}\lesssim\mu_{1,2}$ and 
$\mathcal{O}(10^{-13})\lesssim y_3,y_2$.

\section{Summary}

There are some unsolved problems in the SM. These are, for instance, 
explanations for DM, gauge hierarchy problem, tiny neutrino mass scales, 
baryogenesis, inflation, and the DE. The extended SM without the SUSY, so-called
 NNMSM, could explain the above problems except for the gauge problem by adding 
two adjoint fermions, four vector-like fermions, two gauge singlet real scalars,
 two right-handed neutrinos, and small cosmological constant. In this paper, we 
suggested two types of alternatives (NNMSM-II and NNMSM-III) to the NNMSM by 
reducing additional fields while keeping the above merits of the NNMSM.

First, we have taken a setup that new fermions have a different mass scale of 
new physics. Under the condition, the GCU with the proton stability determines 
the field contents of the NNMSM-II, i.e., two adjoint fermions are added to the 
SM in addition to two gauge singlet real scalars, two right-handed neutrinos, 
and small cosmological constant. The GCU can occur at $\Lambda_{\rm 
GCU}\simeq2.41\times10^{15}$ GeV with two mass scales of new particles as 
$M_3\simeq7.44\times10^9$ GeV and $M_2=300$ GeV. We consider the reheating 
temperature as $T_{RH}\lesssim1.86\times10^8$ GeV in order not to produce the 
stable adjoint fermions in the early universe. This reheating temperature 
requires the following issues. The masses of right-handed neutrino should be 
lighter than $1.86\times10^8$ GeV, so that tiny neutrino mass is realized 
through the Type-I seesaw with relatively small neutrino Yukawa couplings. The 
BAU should be achieved through, for example, the resonant leptogenesis. We have 
also analyzed the stability and triviality conditions by use of recent 
experimental data of Higgs and top masses. We found the parameter regions in 
which the correct abundance of DM can be also realized at the same time. One is 
the lighter $m_S$ region as 63.5GeV$\lesssim m_S\lesssim$64.0 GeV, and the other
 is heavier ones as 708 GeV$\lesssim m_S\lesssim2040$ GeV with the center value 
of top pole mass. Both the regions are almost same as in the NNMSM. This means 
that the differences of runnings of the gauge couplings between NNMSM and 
NNMSM-II does not affect on the stability, triviality, and correct abundance of 
DM in these class of model. Therefore, the favored region for the stability and 
triviality still depends on the top mass rather than runnings of the gauge 
couplings.
The future XENON100 experiment with 20 times 
sensitivity will completely check out the lighter mass region. On the other 
hand, the heavier mass region will also be completely checked by the future 
direct experiments of XENON100$\times$20, XENON1T and/or combined data from 
indirect detections of Fermi$+$CTA$+$Planck at $1\sigma$ CL. 

Second, we have also taken a different setup (NNMSM-III) that includes three 
adjoint fermions ($\lambda_3$ and $\lambda_{2,i}$ ($i=1,2$)) in addition to two 
gauge singlet real scalars and small cosmological constant, but does not have 
right-handed neutrinos. Removing the right-handed neutrino is one of advantages 
of this setup. Then, we have considered two simple cases in this setup. One 
is that all masses of adjoint fermions are the same. The other is that masses 
between $SU(3)_C$ and $SU(2)_L$ adjoint fermions are different. The first and 
the second cases are named as NNMSM-III-A and B, respectively. The GCU can 
occur at  $\Lambda_{\rm GCU}\simeq5.20~(2.77)\times10^{15}$ GeV with 
$M_3=M_{2,i}\simeq2.26\times10^8$ ($M_3\simeq4\times10^9$ and 
$M_{2,i}\simeq6.73\times10^8$) GeV in the NNMSM-III-A (B). Thus, the reheating 
temperature should be $T_{RH}\lesssim5.65\times10^6~(10^8)$ GeV for model A (B).
 The tiny neutrino mass can be realized by two adjoint fermions under $SU(2)_L$ 
through the type-III seesaw mechanism, and the BAU can be achieved, e.g., the 
baryogenesis from the dark sector in both models. We have also investigated 
other initial setups for realizing the GCU in Appendix. These have several 
generations of $\lambda_3$ and/or $\lambda_2$. Regarding the stability, 
triviality, and DM in the NNMSM-III since the differences of the runnings of the
 gauge couplings do not almost affect the stability and triviality bounds, the 
favored regions in the NNMSM-III are also almost the same as in the other 
NNMSMs. Therefore, the NNMSM-III predicts the same mass region of DM and $k$ as 
in the previous models.

Finally, we shortly compare the NNMSMs to any other minimal extensions of the SM
 such as the minimal left-right model (e.g., see~\cite{Marshak:1979fm}), 
neutrino minimal standard model ($\nu$MSM)~\cite{Asaka:2005an}, and some 
supersymmetric extensions. In the context of the left-right model, the GCU, DM, 
tiny neutrino mass, and BAU can be explained but there is not a complete 
analysis for the vacuum stability and triviality with the latest center values 
of the Higgs and top mass. Such discussion might be interesting although the 
RGEs for the Higgs sector are more complicated compared to that in the NNMSMs. 
The $\nu$MSM with three right-handed neutrinos is one simple extension of the 
SM, and can also explain some problems (DM, tiny neutrino mass, and BAU) at the 
same time. In order to achieve the GCU in the $\nu$MSM, some additional 
particles are still needed. In addition, the Higgs sector of the $\nu$MSM is the
 same as in the SM, and thus, the vacuum in the model becomes instable before 
the Planck scale. The degrees of the freedom in the NNMSMs are less than general
 extensions of the SM with the left-right and supersymmetry. These are the 
advantages of the NNMSMs.

\subsection*{Acknowledgement}

We are grateful to Shigeki Matsumoto and Osamu Seto for useful discussions. This
 work is partially supported by Scientific Grant by Ministry of  Education and 
Science, Nos. 00293803, 20244028, 21244036, 23340070, and by the SUHARA Memorial
 Foundation. The works of K.K. and R.T. are supported by Research Fellowships of
 the Japan Society for the Promotion of Science for Young Scientists.

\appendix
\section{Other initial setups for realizing the GCU}

Let us investigate other initial setups for realizing the GCU in this Appendix.

\subsection{{\boldmath $M_{3,i}=M_{2,i}$} case}

Some examples of other initial setups for the GCU with the condition of 
$M_{3,i}=M_{2,i}$ are given in Table~\ref{tab3}. In the table, $N_{\lambda_3}$ 
and $N_{\lambda_2}$ are the number of generations of $\lambda_3$ and 
$\lambda_2$, respectively. We give some comments on those initial setups:
\begin{itemize}
\item The case of $(N_{\lambda_3},N_{\lambda_2})=(1,3)$ can also realize the GCU
 but the case cannot satisfy the constraint from the proton decay. Such case is 
labeled by ``$\ast$" on the number of $N_{\lambda_2}$.

\item The cases of larger number of $N_{\lambda_3}$ needs larger number of 
$N_{\lambda_2}$ for the realization of the GCU. For instance, the GCU can occur 
from  $(N_{\lambda_3},N_{\lambda_2})=(2,\geq3)$. The cases that never realize 
the GCU, e.g. $(N_{\lambda_3},N_{\lambda_2})=(2,1)$, are labeled by 
``$\dagger$''.

\item There are upper bounds on the number of $N_{\lambda_2}$, which come from 
the constraint of the proton decay, for each case of $N_{\lambda_3}$. The cases 
exceeding the corresponding upper bound are also labeled by ``$\ast$'', e.g. 
$(N_{\lambda_3},N_{\lambda_2})=(2,\geq5)$ case.

\item There are some combinations such that the GCU can be realized but the GCU 
scale exceeds the Planck scale, e.g., $(N_{\lambda_3},N_{\lambda_2})=(3,4)$ and 
$(4,5)$ etc. The cases are labeled by ``$\ddagger$''.
\end{itemize}

\subsection{{\boldmath $M_{3,i}\neq M_{2,i}$} case}

Some examples of other initial setups for the GCU allowing $M_{3,i}\neq M_{2,i}$
 are given in Table\ref{tab4}. We give some comments on those initial setups:

\begin{itemize}
\item Upper bounds on $M_{3,i}$ and $M_{2,i}$ are given for each case. 
We take a conservative limit as $\tau\gtrsim10^{34}$ years for the proton decay 
to obtain the upper bounds. Therefore, all cases listed in Tab.\ref{tab4} can 
satisfy the proton decay constraint.

\item $(N_{\lambda_3},N_{\lambda_2})=(1,1)$ case is just the NNMSM-II.

\item $(N_{\lambda_3},N_{\lambda_2})=(1,2)$ case with $M_3=M_{2,i}$ ($M_3\neq M_{2,i}$) is the NNMSM-III-A (B).

\item Larger number of $N_{\lambda_3}$ leads to larger upper bound on $M_{\lambda_{3,i}}$.

\item The cases with $N_{\lambda_{2,i}}=1$ can satisfy the constraint from the 
proton decay but these are excluded by the LHC experiment searching for the 
$SU(2)_L$ triplet particle. The cases are labeled by ``$\star$''.
\end{itemize}
Hierarchical mass spectra for $M_{3,i}$ and $M_{2,i}$ such as $M_{3,1}<M_{3,2}$ 
might also realize the GCU but we do not consider the hierarchical mass spectra 
case for minimality of the models in this work.

\begin{table}
\begin{center}
\begin{tabular}{c|c|c|c|c|c}
\hline
$N_{\lambda_3}$ & $N_{\lambda_2}$ & $M_{NP}$ [GeV] & $\Lambda_{\rm GCU}$ [$10^{15}$ GeV] & $\alpha_{\rm GCU}^{-1}$ & $\tau$ [$10^{33}$ years] \\
\hline\hline
1 & 3$^\ast$ & $2.21\times10^{11}$ & 1.41 & 39.3 & $<8.2\times10^{33}$ years 
\\
\hline
2 & 1,2$^\dagger$ & - & - & - & - \\
\cline{2-6}
 & 3 & $5.80\times10^9$ & 99.2 & 36.4 & $2.88_{-1.19}^{+3.06}\times10^7$ \\
\cline{2-6}
 & 4 & $1.09\times10^{12}$ & 5.20 & 38.3 & $1.89_{-0.78}^{+2.00}\times10^2$ \\
\cline{2-6}
 & $\geq5^\ast$ & $\geq6.47\times10^{12}$ & $\leq1.90$ & $\geq39.0$ & $<8.2\times10^{33}$ years \\
\hline
3 & 1-3$^\dagger$ & - & - & - & - \\
\cline{2-6}
 & 4$^\ddagger$ & $1.33\times10^{11}$ & $>M_{\rm pl}$ & 34.6 & $>\mathcal{O}(10^{35})$ years \\
\cline{2-6}
 & 5,6 & $(0.52-1.83)\times10^{13}$ & $(5.20-23.0)$ & $(37.3-38.3)$ & $>\mathcal{O}(10^{35})$ years \\
\cline{2-6}
 & 7 & $3.47\times10^{13}$ & $2.45$ & $38.8$ & $8.94_{-3.68}^{+9.49}$ \\
\cline{2-6}
 & $\geq8^\ast$ & $\geq5.11\times10^{13}$ & $\leq1.55$ & $\geq39.1$ & $<8.2\times10^{33}$ years \\
\hline
4 & 1-4$^\ast$ & - & - & - & - \\
\cline{2-6}
 & 5$^\ddagger$ & $2.74\times10^{12}$ & $>M_{\rm pl}$ & 32.8 & $>\mathcal{O}(10^{35})$ years \\
\cline{2-6}
 & 6-8 & $(2.39-7.51)\times10^{13}$ & $(5.20-99.2)$ & $(36.4-38.3)$ & $>\mathcal{O}(10^{35})$ years \\
\cline{2-6}
 & 9 & $9.49\times10^{13}$ & $2.85$ & $38.7$ & $16.4_{-6.75}^{+17.4}$ \\
\cline{2-6}
 & $\geq10^\dagger$ & $\geq1.11\times10^{14}$ & $\leq1.90$ & $\geq39.0$ & $<8.2\times10^{33}$ years \\
\hline
5 & 1-5$^\ast$ & - & - & - & - \\
\cline{2-6}
 & 6$^\ddagger$ & $5.11\times10^{12}$ & $>M_{\rm pl}$ & 31.0 & $>\mathcal{O}(10^{35})$ years \\
\cline{2-6}
 & 7-10 & $(1.75-7.51)\times10^{14}$ & $(5.20-417)$ & $(35.5-38.3)$ & $>\mathcal{O}(10^{35})$ years \\
\cline{2-6}
 & 11 & $1.85\times10^{14}$ & $3.15$ & $38.7$ & $24.6_{-10.1}^{+26.1}$ \\
\cline{2-6}
 & 12 & $1.93\times10^{14}$ & $2.20$ & $38.9$ & $5.76_{-2.37}^{+6.61}$ \\
\cline{2-6}
 & $\geq13^\dagger$ & $\geq1.99\times10^{14}$ & $\leq1.68$ & $\geq39.1$ & $<8.2\times10^{33}$ years \\
\hline
6 & 1-6$^\ast$ & - & - & - & - \\
\cline{2-6}
 & 7,8$^\ddagger$ & $(4.76-8.67)\times10^{14}$ & $>M_{\rm pl}$ & (29.3-34.6) & $>\mathcal{O}(10^{35})$ years \\
\cline{2-6}
 & 9-12 & $(3.09-3.85)\times10^{14}$ & $(5.20-99.2)$ & $(36.4-38.3)$ & $>\mathcal{O}(10^{35})$ years \\
\cline{2-6}
 & 13 & $2.99\times10^{14}$ & $3.39$ & $38.6$ & $32.8_{-13.5}^{+34.9}$ \\
\cline{2-6}
 & 14 & $2.92\times10^{14}$ & $2.45$ & $38.8$ & $8.90_{-3.66}^{+9.45}$ \\
\cline{2-6}
 & $\geq15^\dagger$ & $\geq2.86\times10^{14}$ & $\leq1.90$ & $\geq39.0$ & $<8.2\times10^{33}$ years \\
\hline
7 & 1-7$^\ast$ & - & - & - & - \\
\cline{2-6}
 & 8,9$^\ddagger$ & $(0.204-1.34)\times10^{16}$ & $>M_{\rm pl}$ & $(27.7-33.7)$ & $>\mathcal{O}(10^{35})$ years \\
\cline{2-6}
 & 10-14 & $(0.462-1.04)\times10^{15}$ & $(5.20-259)$ & $(35.8-38.3)$ & $>\mathcal{O}(10^{35})$ years \\
\cline{2-6}
 & 15 & $4.27\times10^{14}$ & $3.57$ & $38.6$ & $40.8_{-16.8}^{+43.3}$ \\
\cline{2-6}
 & 16 & $4.02\times10^{14}$ & $2.66$ & $38.8$ & $12.5_{-5.14}^{+13.3}$ \\
\cline{2-6}
 & 17 & $3.83\times10^{14}$ & $2.10$ & $38.9$ & $4.83_{-1.99}^{+5.13}$ \\
\cline{2-6}
 & $\geq18^\dagger$ & $\leq3.68\times10^{14}$ & $\leq1.74$ & $\geq39.1$ & $<8.2\times10^{33}$ years \\
\hline
\end{tabular}
\end{center}
\caption{Examples for realizing the GCU with several generations of new adjoint 
fermions under the condition of $M_{NP}=M_{3,i}=M_{2,i}$ in the NNMSM-III. The 
label ``$\ast$'' on the number of $N_{\lambda_2}$ means that the cases cannot 
satisfy the proton decay constraint. The label ``$\dagger$'' means that the 
cases never realize the GCU. The label ``$\ddagger$'' means that the cases can 
realize the GCU but the scale is higher than the Planck scale. The cases without
 any labels can realize the GCU at an energy scale lower than the Planck scale 
and satisfy the proton decay constraint.}
\label{tab3}
\end{table}
\begin{table}
\begin{center}
\begin{tabular}{c|c|c|c}
\hline
$N_{\lambda_3}(M_{3,i})$ & $N_{\lambda_2}(M_{2,i})$ & $\Lambda_{\rm GCU}$ [$10^{15}$ GeV] & $\alpha_{\rm GCU}^{-1}$ \\
\hline\hline
1 ($M_3\lesssim4\times10^9$ GeV) & 3 ($M_{2,i}\lesssim1.08\times10^{11}$ GeV) & $2.77$ & 38.8 \\
\cline{2-2}
 & 4 ($M_{2,i}\lesssim1.36\times10^{12}$ GeV) & & \\
\cline{2-2}
 & 5 ($M_{2,i}\lesssim6.26\times10^{12}$ GeV) & & \\
\cline{2-2}
 & 6 ($M_{2,i}\lesssim1.73\times10^{13}$ GeV) & & \\
\cline{2-2}
 & 7 ($M_{2,i}\lesssim3.57\times10^{13}$ GeV) & & \\
\cline{2-2}
 & 8 ($M_{2,i}\lesssim6.14\times10^{13}$ GeV) & & \\
\cline{2-2}
 & 9 ($M_{2,i}\lesssim9.39\times10^{13}$ GeV) & & \\
\hline
2  ($M_{3,i}\lesssim3\times10^{12}$ GeV) & 1$^\star$ ($M_{2,i}\lesssim126$ GeV) & 2.91 & 38.7 \\
\cline{2-2}
 & 2 ($M_{2,i}\lesssim6.08\times10^8$ GeV) & & \\
\cline{2-2}
 & 3 ($M_{2,i}\lesssim1.03\times10^{11}$ GeV) & & \\
\cline{2-2}
 & 4 ($M_{2,i}\lesssim1.34\times10^{12}$ GeV) & & \\
\cline{2-2}
 & 5 ($M_{2,i}\lesssim6.22\times10^{12}$ GeV) & & \\
\cline{2-2}
 & 6 ($M_{2,i}\lesssim1.73\times10^{13}$ GeV) & & \\
\cline{2-2}
 & 7 ($M_{2,i}\lesssim3.61\times10^{13}$ GeV) & & \\
\cline{2-2}
 & 8 ($M_{2,i}\lesssim6.26\times10^{13}$ GeV) & & \\
\cline{2-2}
 & 9 ($M_{2,i}\lesssim9.60\times10^{13}$ GeV) & & \\
\cline{1-2}
3 ($M_{3,i}\lesssim3\times10^{13}$ GeV) & 1$^\star$ ($M_{2,i}\lesssim124$ GeV) & & \\
\cline{2-2}
 & 2 ($M_{2,i}\lesssim6.05\times10^8$ GeV) & & \\
\cline{2-2}
 & 3 ($M_{2,i}\lesssim1.03\times10^{11}$ GeV) & & \\
\cline{2-2}
 & 4 ($M_{2,i}\lesssim1.33\times10^{12}$ GeV) & & \\
\cline{2-2}
 & 5 ($M_{2,i}\lesssim6.22\times10^{12}$ GeV) & & \\
\cline{2-2}
 & 6 ($M_{2,i}\lesssim1.74\times10^{13}$ GeV) & & \\
\cline{2-2}
 & 7 ($M_{2,i}\lesssim3.62\times10^{13}$ GeV) & & \\
\cline{2-2}
 & 8 ($M_{2,i}\lesssim6.27\times10^{13}$ GeV) & & \\
\cline{2-2}
 & 9 ($M_{2,i}\lesssim9.61\times10^{13}$ GeV) & & \\
\hline
4 ($M_{3,i}\lesssim9\times10^{13}$ GeV) & 1$^\star$ ($M_{2,i}\lesssim78.3$ GeV) & 3.27 & 38.6 \\
\cline{2-2}
 & 2 ($M_{2,i}\lesssim5.06\times10^8$ GeV) & & \\
\cline{2-2}
 & 3 ($M_{2,i}\lesssim9.42\times10^{10}$ GeV) & & \\
\cline{2-2}
 & 4 ($M_{2,i}\lesssim1.28\times10^{12}$ GeV) & & \\
\cline{2-2}
 & 5 ($M_{2,i}\lesssim6.16\times10^{12}$ GeV) & & \\
\cline{2-2}
 & 6 ($M_{2,i}\lesssim1.75\times10^{13}$ GeV) & & \\
\cline{2-2}
 & 7 ($M_{2,i}\lesssim3.70\times10^{13}$ GeV) & & \\
\cline{2-2}
 & 8 ($M_{2,i}\lesssim6.48\times10^{13}$ GeV) & & \\
\cline{2-2}
 & 9 ($M_{2,i}\lesssim1.00\times10^{14}$ GeV) & & \\
\hline
\end{tabular}
\end{center}
\caption{Examples for realizing the GCU with several generations of new adjoint 
fermions and the condition of $M_{3,i}\neq M_{2,i}$ in the NNMSM-III. All 
cases listed in this table can satisfy the proton decay constraint but the cases with ``$\star$'' are excluded by the LHC experiment searching for the $SU(2)_L$ adjoint fermion.}
\label{tab4}
\end{table}

\end{document}